\documentclass[a4paper,11pt]{article}
\usepackage[dvips]{graphicx}
\usepackage{amssymb}
\usepackage{amsmath}

\usepackage{cite}

\begin{document}

\title{F-theory Yukawa Couplings and Supersymmetric Quantum Mechanics}
\author{V.K.Oikonomou\thanks{
voiko@physics.auth.gr}\\
Technological Education Institute of Serres, \\
Department of Informatics and Communications 62124 Serres, Greece\\
and\\
Department of Theoretical Physics Aristotle University of Thessaloniki,\\
Thessaloniki 541 24 Greece} \maketitle

\begin{abstract}
The localized fermions on the intersection curve $\Sigma$ of
D7-branes, are connected to a $N=2$ supersymmetric quantum
mechanics algebra. Due to this algebra the fields obey a global
$U(1)$ symmetry. This symmetry restricts the proton decay
operators and the neutrino mass terms. Particularly, we find that
several proton decay operators are forbidden and the Majorana mass
term is the only one allowed in the theory. A special SUSY QM
algebra is studied at the end of the paper. In addition we study
the impact of a non-trivial holomorphic metric perturbation on the
localized solutions along each matter curve. Moreover, we study
the connection of the localized solutions to an $N=2$
supersymmetric quantum mechanics algebra when background fluxes
are turned on.
\end{abstract}

\section*{Introduction}

F-theory
\cite{vafa1,origins1,origins2,vafa2,vafa3,review1,review2,review3,review4,review5,uplift,f1,f2,f3,f4,f5,f6,f7,f8,f9,f10,f11,f12,f13,f14,f15,f16,f17,f18,f19,f21,f22,f23,f24,f25,f26,f27,f28,f29,f30,f31,f32,f33,f34,f35,f36,f37,f38,f39,f40,f41,f42,f43,f44,f45,f46,f47,f48,f49,f50,f51,f52,f53,f54,f55,f56,f57,f57a,f58b,f58,f59,f60,f61,f62,f63,f64,f65,f66,f67,f68,f69,f70,f71,f72},
has received a prominent role lately, due to the fact that GUTs
can be consistently constructed and well founded, within
F-theory's wide theoretical framework. It is an 12-dimensional
theory that consists of toroidal elliptic fibrations over
Calabi-Yau manifolds. D7 branes are essential to the theory, since
the D7 branes are located on the $T^2$ fiber. The modulus of the
torus is a varying parameter and is related to the axio-dilaton.
Thereupon, we can say that F-theory is an UV completion of type
IIB superstring theory with 7-branes. For comprehensive reviews on
the formulation of F-theory GUT's see
\cite{review1,review2,review3,review4,review5,f9,f12}

\noindent One of it's most interesting outcomes, is that within
F-theory we can produce many phenomenological features of GUTs
with gravity being excluded from the theoretical apparatus (for
recent work on realistic F-theory GUTs models see
\cite{vafa2,vafa3,review1,review2,review3,review4,uplift,f1,f2,f3,f4,f5,f6,f7,f8,f9,f10,f11,f12,f13,f14,f15,f16,f17,f18,f19,f21,f22,f23,f24,f25,f26,f27,f28,f29,f30,f31,f32,f33,f34,f35,f36,f37,f38,f39,f40,f41,f42,f43,f44,f45,f46,f47,f48,f49,f50,f51,f52,f53,f54,f55,f56,f57,f58,f59,f60,f61}).
Moreover, certain phenomenological features that was not possible
to be realized within the perturbative framework of superstring
theories, for example the couplings $5_H\times 10_M\times 10_M$
for $SU(5)$, or the spinor $16$ representation of $SO(10)$, now
can be consistently incorporated to the phenomenological outcomes
of the theory.

\noindent Complex manifolds singularities play a critical role in
F-theory phenomenology
\cite{vafa1,vafa2,review1,review2,review3,review4,review5,uplift,f1},
with gauge groups realized by the geometry of singularities.
Additionally, $N=1$, $D=4$ supersymmetric gauge theories arise
when F-theory is compactified on Calabi-Yau fourfolds
\cite{origins2,vafa2,review4,f1}.

\noindent One of the most important features that a UV completion
of the Standard Model must somehow explain is the hierarchical
structure of fermion masses and mixings. In F-theory GUTs much
work has been done towards this direction
\cite{vafa2,f1,f5,f13,f16,f28,f39,f67,f68} and also in order to
explain the neutrino sector, see \cite{f69} and also
\cite{f39,f41,f68,f70,f71}. Yukawa couplings in F-theory are
obtained by calculating overlapping integrals of three matter
curves wave functions over a complex surface $S$. Hence Yukawas
depend drastically on the local structure of the theory, near the
intersection point of the three matter curves (nonetheless the
global structure of the theory affects the normalization of the
wave functions).

\noindent In this work we shall consider the localized fields that
are generated on the intersection curve $\Sigma$ of D7-branes
without external gauge fluxes, and also the Yukawa couplings
generated by the intersection of three matter curves. The fields
that have localized solutions along the matter curve $\Sigma$, are
connected \cite{oikonomouf} to an $N=2$ supersymmetric quantum
mechanics algebra \cite{susyqm,susyqm1} and the number of the zero
modes is connected to the Witten index of the susy algebra. We
find that every localized field on the matter curve obeys a hidden
global $U(1)$ symmetry. We shall require that this symmetry holds
even at the intersection point of three matter curves. The
conditions that must hold in order this to happen, pose some
restrictions on various proton decay operators and on the
operators that give masses to the neutrinos. Furthermore, we study
the impact of a certain type of susy quantum algebra on the Yukawa
coupling that gives mass to the top quark. The results are
interesting, since, the imposed conditions result to a form of
wave functions with de-localized Higgs. Moreover, we shall include
the effects of a gravitational backreaction on the complex surface
$S$, in terms of linear perturbations of the Euclidean metric. We
conclude that the spectral problems of the perturbed and
unperturbed system are identical, due to the topological
invariance of the index of the corresponding operators. We shall
examine all the matter curves. Finally, we shall check whether we
can relate a SUSY QM algebra to the fermions localized along the
three matter curves, in the case we introduce constant background
gauge fluxes.

This paper is organized as follows: In section 1 we describe in
brief the F-theory setup we shall use, that is, D7-branes
intersections, matter curves and the eight dimensional
Super-Yang-Mills (SYM) theory. In section 2 we give in short, a
self-contained review of the supersymmetric quantum mechanics
algebra. In section 3 we connect the localized solutions of the
BPS equations of motion, to an $N=2$ supersymmetric quantum
mechanics algebra and also we study the impact of a non-trivial
linear perturbation of the metric on the localized solutions. In
section 4 we examine the localized solutions of the fermionic
system under the influence of background gauge fluxes. In section
5 we study the impact of a certain type of susy quantum algebra on
the top quark Yukawa coupling. In section 6 we study the $U(1)$
symmetries and the restrictions these imply to the proton decay
operators and to the neutrino mass operators. Finally in section 7
we present the conclusions.

\section{Localized Fermions on D7-Branes Intersections}

\noindent We shall consider F-theory compactifications on a
Calabi-Yau fourfold. This manifold is an elliptic $K3$ fibration
over a complex dimension two surface $S$. Locally the theory can
be described by the worldvolume of an ADE type D7 brane wrapping
\linebreak $R^{1,3}\times S$ over the Calabi-Yau fourfold. The
resulting $d=4$ theory is an $N=1$ supersymmetric theory
\cite{vafa2,f1}. Our analysis is based mostly on references
\cite{vafa2,f1}.

\noindent The physics of the D7-branes wrapping $S$ can be
described in terms of an $D=8$ twisted Super Yang-Mills on
$R^{3,1}\times S$. The supersymmetric multiplets contain the gauge
field plus a complex scalar $\varphi$ and the set of adjoint
fermions $\eta ,\psi , \chi $. We parameterize the complex surface
$S$ using the local coordinates ($z_1,z_2$). Then the
supermultiplets are:
\begin{equation}\label{gauge1}
A=A_{\mu}\mathrm{d}x^{\mu}+A_{m}\mathrm{d}z^{m}+A_{\bar{m}}\mathrm{d}\bar{z}^{m},{\,}{\,}{\,}{\,}\varphi
=\varphi_{12}{\,}\mathrm{d}z^1\wedge \mathrm{d}z^2
\end{equation}
and additionally,
\begin{equation}\label{gauge2}
\psi_{a}=\psi_{a\bar{1}}\mathrm{d}\bar{z}^1+\psi_{a\bar{2}}\mathrm{d}\bar{z}^2,{\,}{\,}{\,}{\,}\chi_a
=\chi_{a12}{\,}\mathrm{d}z^1\wedge \mathrm{d}z^2
\end{equation}
with $a=1,2$ and $m=1,2$.

\noindent The gauge multiplet $(A_{\mu},\eta)$ together with the
chiral multiplets $(A_{\bar{m}},\psi_{\bar{m}})$ and
$(\varphi_{12},\chi_{12})$, plus their complex conjugates
constitute the $N=1$, $D=4$ supersymmetric theory. Omitting the
kinetic terms (we shall use the kinetic terms later on in this
article), the bilinear in fermions  part of the action is,
\begin{equation}\label{bilinear}
I_F=\int_{R^{1,3}\times S}\mathrm{d}x^4\mathrm{Tr}\Big{(}\chi
\wedge\partial_A\psi+2{\,}i\sqrt{2}{\,}\omega \wedge\partial_A\eta
\wedge \psi+\frac{1}{2}\psi \wedge [\varphi,\psi]+\sqrt{2}{\,}\eta
[\bar{\varphi},\chi]+\mathrm{h.c.}\Big{)}
\end{equation}
with $\omega$ is the fundamental K\"{a}hler form of the complex
surface $S$. The variation of $\eta$, $\psi$ and $\chi$, yields
the equations of motion \cite{vafa2,f1}:
\begin{align}
& \omega\wedge \partial_A \psi+\frac{i}{2}[\bar{\phi},\chi]=0 \\
\notag & \bar{\partial}_A\chi-2i\sqrt{2}\omega\wedge \partial
\eta-[\varphi,\psi]=0 \\ \notag &
\bar{\partial}_A\psi-\sqrt{2}{\,}[\bar{\varphi},\eta]=0
\end{align}

\noindent Before we proceed in details on how to find zero modes,
we review in brief some issues, regarding the supersymmetric
quantum mechanics algebra which we shall frequently use in the
subsequent sections.

\section{$N=2$ Supersymmetric Quantum Mechanics Algebra}

Consider a quantum system, described by a Hamiltonian $H$ and
characterized by the set $\{H,Q_1,...,Q_N\}$, with $Q_i$
self-adjoint operators. The quantum system is called
supersymmetric, if,
\begin{equation}\label{susy1}
\{Q_i,Q_j\}=H\delta_{i{\,}j}
\end{equation}
with $i=1,2,...N$. The $Q_i$  are the supercharges and the
Hamiltonian ``$H$" is called SUSY Hamiltonian. The algebra
(\ref{susy1}) describes the N-extended supersymmetry with zero
central charge. Owing to the anti-commutativity, the Hamiltonian
can be written as,
\begin{equation}\label{susy3}
H=2Q_1^2=Q_2^2=\ldots =2Q_N^2=\frac{2}{N}\sum_{i=1}^{N}Q_i^2.
\end{equation}
A supersymmetric quantum system $\{H,Q_1,...,Q_N\}$ is said to
have unbroken supersymmetry, if its ground state vanishes, that is
$E_0=0$. In the case $E_0>0$, that is, for a positive ground state
energy, susy is said to be broken.

\noindent In order supersymmetry is unbroken, the Hilbert space
eigenstates must be annihilated by the supercharges,
\begin{equation}\label{s1}
Q_i |\psi_0^j\rangle=0
\end{equation}
for all $i,j$.

\noindent The $N=2$ algebra (``$N=2$ SUSY QM", or ``SUSY QM"
thereafter) consists of two supercharges $Q_1$ and $Q_2$ and a
Hamiltonian $H$, which obey the following,
\begin{equation}\label{sxer2}
\{Q_1,Q_2\}=0,{\,}{\,}{\,}H=2Q_1^2=2Q_2^2=Q_1^2+Q_2^2
\end{equation}
We use the complex supercharge $Q$ and it's adjoint $Q^{\dag}$
defined as,
\begin{equation}\label{s2}
Q=\frac{1}{\sqrt{2}}(Q_{1}+iQ_{2}){\,}{\,}{\,}{\,}{\,}{\,}{\,}{\,}{\,}Q^{\dag}=\frac{1}{\sqrt{2}}(Q_{1}-iQ_{2})
\end{equation}
which satisfy the following equations,
\begin{equation}\label{s23}
Q^{2}={Q^{\dag}}^2=0
\end{equation}
and also are related to the Hamiltonian as,
\begin{equation}\label{s4}
\{Q,Q^{\dag}\}=H
\end{equation}
A very important operator that is inherent to the definition of a
SUSY QM system is the Witten parity, $W$, which, for a $N=2$
algebra, is defined as,
\begin{equation}\label{s45}
\{W,Q\}=\{W,Q^{\dag}\}=0,{\,}{\,}{\,}{\,}{\,}{\,}{\,}{\,}{\,}{\,}[W,H]=0
\end{equation}
and satisfies,
\begin{equation}\label{s5}
W^{2}=I
\end{equation}
The main and important use of the operator $W$ is that, by using
it, we can span the Hilbert space $\mathcal{H}$ of the quantum
system to positive and negative Witten parity spaces, defined as,
$\mathcal{H}^{\pm}=P^{\pm}\mathcal{H}=\{|\psi\rangle :
W|\psi\rangle=\pm |\psi\rangle\} $. Therefore, the quantum system
Hilbert space $\mathcal{H}$ is decomposed into the eigenspaces of
$W$, hence $\mathcal{H}=\mathcal{H}^+\oplus \mathcal{H}^-$.  Since
each operator acting on the vectors of $\mathcal{H}$ can be
represented by $2N\times 2N$ matrices, we use the representation:
\begin{equation}\label{s7345}
W=\bigg{(}\begin{array}{ccc}
  I & 0 \\
  0 & -I  \\
\end{array}\bigg{)}
\end{equation}
with $I$ the $N\times N$ identity matrix. Recalling that $Q^2=0$
and $\{Q,W\}=0$, the supercharges take the form,
\begin{equation}\label{s7}
Q=\bigg{(}\begin{array}{ccc}
  0 & A \\
  0 & 0  \\
\end{array}\bigg{)},{\,}{\,}{\,}{\,}{\,}{\,}{\,}{\,}{\,}Q^{\dag}=\bigg{(}\begin{array}{ccc}
  0 & 0 \\
  A^{\dag} & 0  \\
\end{array}\bigg{)}
\end{equation}
which imply,
\begin{equation}\label{s89}
Q_1=\frac{1}{\sqrt{2}}\bigg{(}\begin{array}{ccc}
  0 & A \\
  A^{\dag} & 0  \\
\end{array}\bigg{)},{\,}{\,}{\,}{\,}{\,}{\,}{\,}{\,}{\,}Q_2=\frac{i}{\sqrt{2}}\bigg{(}\begin{array}{ccc}
  0 & -A \\
  A^{\dag} & 0  \\
\end{array}\bigg{)}
\end{equation}
The $N\times N$ matrices $A$ and $A^{\dag}$, are generalized
annihilation and creation operators with $A$ acting as $A:
\mathcal{H}^-\rightarrow \mathcal{H}^+$ and $A^{\dag}$ as,
$A^{\dag}: \mathcal{H}^+\rightarrow \mathcal{H}^-$. In the
representation (\ref{s7345}), (\ref{s7}), (\ref{s89}) the quantum
mechanical Hamiltonian $H$, can be cast in a diagonal form,
\begin{equation}\label{s11}
H=\bigg{(}\begin{array}{ccc}
  AA^{\dag} & 0 \\
  0 & A^{\dag}A  \\
\end{array}\bigg{)}
\end{equation}
We denote $n_{\pm}$ the number of zero modes of $H_{\pm}$. The
Witten index for Fredholm operators is defined as,
\begin{equation}\label{phil}
\Delta =n_{-}-n_{+}
\end{equation}
When the Witten index is non-zero integer, supersymmetry is
unbroken and in the case the Witten index is zero, if
$n_{+}=n_{-}=0$ supersymmetry is broken, while if $n_{+}=
n_{-}\neq 0$ supersymmetry is unbroken.

\noindent The Fredholm index of the operator $A$ and the Witten
index are related as,
\begin{align}\label{ker1}
&\Delta=\mathrm{ind} A=\mathrm{dim}{\,}\mathrm{ker}
A-\mathrm{dim}{\,}\mathrm{ker} A^{\dag}=
\\ \notag &
\mathrm{dim}{\,}\mathrm{ker}A^{\dag}A-\mathrm{dim}{\,}\mathrm{ker}AA^{\dag}=\mathrm{dim}{\,}\mathrm{ker}
H_{-}-\mathrm{dim}{\,}\mathrm{ker} H_{+}
\end{align}
We shall consider only Fredholm operators.

\section{Intersecting Matter Curves, Localized Fermions and Supersymmetric Quantum Mechanics}

The localized fields on each matter curve on $S$ are related to a
SUSY QM algebra, as shown in \cite{oikonomouf}. In order to make
this article self contained we review the basic facts (for details
see \cite{oikonomouf}). The localized fermion fields exist on a
matter curve $\Sigma$ which is the intersection of the complex
surfaces $S$ and $S'$. In order to preserve $N=1$ supersymmetry in
$D=4$, the theory defined on $R^{1,3}\times \Sigma$ must be $D=6$
twisted super Yang-Mills \cite{vafa2,f1}.

\noindent Solving the $D=8$ equations of motion for the twisted
fermions we find how localized fermion matter on $\Sigma$ results
from zero modes of the $D=8$ bulk theory. Consider three matter
curves denoted as $\Sigma_i$, with $i=1,2,3$. Each matter curve
has a group $G_i$, that on the intersection point further enhances
to a higher group $G_p$. A non-trivial background for the adjoint
scalar is required in order to extract the localized fermionic
solutions of the eight dimensional theory on $S$ \cite{vafa2,f1},
which is equal to \cite{vafa2,f1}:
\begin{equation}\label{backscalar1}
\langle \varphi \rangle = m^2z_1Q_1+m^2z_2Q_2
\end{equation}
In the above, $Q_1$ and $Q_2$ are the $U(1)$ generators that are
included in the enhancement group $G_p$ at the intersection point,
and ``$m_1$" and ``$m_2$" are mass scales related to the F-theory
scale $M_*$. Taking $m_1=m_2=m$ will simplify things but will not
change the results.

\noindent The three matter curves can intersect at a point which
is $(z_1,z_2)=(0,0)$. The adjoint vacuum expectation value
(\ref{backscalar1}) resolves the $G_p$ singularity at the
intersection point. The three different curves $\Sigma_1$,
$\Sigma_2$, $\Sigma_3$ are defined by the loci $z_1=0$, $z_2=0$
and $z_1+z_2=0$ respectively. Note that each curve represents a
fermion under the $U(1)$ charges, the curves can be classified
according to the table,
\begin{center}\label{table2}
\begin{tabular}{|c|c|c|}
  \hline
  \bf{matter curve} & ($q_1,q_2$) & \bf{surface locus} \\
  \hline
  $\Sigma_1$ & $(q_1,0)$  & $z_1=0$\\
  \hline
  $\Sigma_2$ & $(0,q_2)$ & $z_2=0$\\
  \hline
  $\Sigma_3$ & $(-q_1,-q_2)$ & $z_1+z_2${\,}={\,}0\\
  \hline
\end{tabular}
\\ \bigskip{ \bfseries{Table 1: Charge Classification of the three matter curves}}
\end{center}
\bigskip
We assume that the K\"{a}hler form of $S$ is the canonical form,
\begin{equation}\label{kaeheregut}
\omega =\frac{i}{2}\large{(}\mathrm{d}z^1\wedge
\mathrm{d}\bar{z}^1+\mathrm{d}z^2\wedge \mathrm{d}\bar{z}^2\large{)}
\end{equation}
The coordinates $z_1$ and $z_2$ that parameterize $S$, describe
the intersection $\Sigma$ in transverse and tangent directions
respectively. With $\omega$ as in (\ref{kaeheregut}) and
neglecting the $z_2$ derivatives, the equations of motion can be
written as \cite{vafa2,f1}:
\begin{align}\label{eqmotion1}
& \sqrt{2}\partial_1\eta
-m^2z_1q_1\psi_{\bar{2}}=0{\,}{\,}{\,}{\,}{\,}{\,}{\,}{\,}{\,}{\,}{\,}{\,}{\,}{\,}{\,}{\,}{\,}{\,}\partial_1
\psi_{\bar{1}}-m^2\bar{z}_1q_1\chi =0
\\  \notag & \partial_1\psi_{\bar{2}}-\sqrt{2}{\,}m^2z_1q_1\eta
=0
{\,}{\,}{\,}{\,}{\,}{\,}{\,}{\,}{\,}{\,}{\,}{\,}{\,}{\,}{\,}{\,}{\,}{\,}\bar{\partial}
        _1\chi-m^2z_1q_1\psi_{\bar{1}}=0
\end{align}
where $(q_1,q_2)$ are the $U(1)$ charges of the fermions belonging
to an irreducible representation $(R,q_1,q_2)$ of $G_S\times
U(1)_1\times U(1)_2$ (note that $Q_1$ is the $U(1)_1$ generator
and $Q_2$ is the $U(1)_2$ generator). Taking the adjoint vacuum
expectation value (\ref{backscalar1}) the equations of motion can
be cast as:
\begin{align}\label{eqmotion2}
& \partial_2\psi_{\bar{2}}+\partial_1\psi_{\bar{1}}
-m^2(\bar{z}_1q_1+\bar{z}_2q_2)\chi =0
\\  \notag & {\,}{\,}{\,}{\,}{\,}{\,}{\,}{\,}{\,}{\,}{\,}{\,}{\,}{\,}{\,}{\,}{\,}{\,}{\,}\bar{\partial}_1\chi-m^2(z_1q_1+z_2q_2)\psi_{\bar{1}} =0
\\  \notag & {\,}{\,}{\,}{\,}{\,}{\,}{\,}{\,}{\,}{\,}{\,}{\,}{\,}{\,}{\,}{\,}{\,}{\,}{\,}\bar{\partial}_2\chi-m^2(z_1q_1+z_2q_2)\psi_{\bar{2}} =0
\end{align}

\subsection{Localized fermion around $z_1=0$}

The curve $\Sigma_1$, corresponds to $q_2=0$. The fermions
localized at $z_1=0$ are obtained by (\ref{eqmotion2}) and are
equal to \cite{f1}:
\begin{equation}\label{locali1}
\psi_{\bar{2}}=0,{\,}{\,}{\,}{\,}{\,}{\,}\chi_1=f(z_2)e^{-q_1m^2|z_1|^2},{\,}{\,}{\,}\psi_{\bar{1}}=-\chi.
\end{equation}
with $f(z_2)$ a $z_2$-dependent holomorphic function. We can
connect a $N=2$ SUSY QM algebra to this matter curve. Indeed, we
can define the matrix $D_1$ and also $D^{\dag}_1$ as follows,
\begin{equation}\label{dmatrix1}
D_1=\left(%
\begin{array}{cc}
  \partial_1 & -m^2\bar{z}_1q_1 \\
  -m^2z_1q_1 & \bar{\partial}_{1} \\
\end{array}%
\right)
\end{equation}
and,
\begin{equation}\label{dmatrix23}
D^{\dag}_1=\left(%
\begin{array}{cc}
  \bar{\partial}_{1} & -m^2\bar{z}_1q_1 \\
  -m^2z_1q_1 & \partial_{1} \\
\end{array}%
\right)
\end{equation}
acting on,
\begin{equation}\label{wee33}
\left(%
\begin{array}{c}
  \psi_{\bar{1}} \\
  \chi_1 \\
\end{array}%
\right)
\end{equation}
The solutions of the equations of motion (\ref{eqmotion2}) with
$\psi_{\bar{1}}$ and $\chi_{\bar{1}}$ the zero modes of $D_1$. The
Fredholm index $I_D$, of the operator $D_1$, is equal to,
\begin{equation}\label{indexd}
\mathrm{ind}\mathrm{I}_{D_1}=\mathrm{dim{\,}ker}(D_1^{\dag})-\mathrm{dim{\,}ker}(D_1)
\end{equation}
which is equal to the number of zero modes of $\mathcal{D}_1$
minus the number of zero modes of $\mathcal{D}_1^{\dag}$.

\noindent Using $D_1$ we can define the $N=2$ supersymmetric
quantum mechanical system by defining the supercharges $Q$ and
$Q^{\dag}$,
\begin{equation}\label{wit2}
Q=\bigg{(}\begin{array}{ccc}
  0 & D_1 \\
  0 & 0  \\
\end{array}\bigg{)}{\,}{\,}{\,}{\,}{\,}{\,}{\,}{\,}Q^{\dag}=\bigg{(}\begin{array}{ccc}
  0 & 0 \\
  D_1^{\dag} & 0  \\
\end{array}\bigg{)}
\end{equation}
Also the Hamiltonian of the system can be written,
\begin{equation}\label{wit4}
H=\bigg{(}\begin{array}{ccc}
  D_1D_1^{\dag} & 0 \\
  0 & D_1^{\dag}D_1  \\
\end{array}\bigg{)}
\end{equation}
The above matrices obey, $\{Q,Q^{\dag}\}=H$, $Q^2=0$,
${Q^{\dag}}^2=0$. Like so, the Witten index of the $N=2$
supersymmetric quantum mechanics system, is related to the index
$I_{D_1}$ of the operator $D_1$. Indeed we have $I_{D_1}=-\Delta$,
because,
\begin{equation}\label{ker}
I_{D_1}=\mathrm{dim}{\,}\mathrm{ker}
D_1^{\dag}-\mathrm{dim}{\,}\mathrm{ker} D_1=
\mathrm{dim}{\,}\mathrm{ker}D_1D_1^{\dag}-\mathrm{dim}{\,}\mathrm{ker}D_1^{\dag}D_1=-\mathrm{ind}D_1=-\Delta=n_--n_+
\end{equation}
with $n_{-}$ and $n_+$ defined in the previous section.
Accordingly, the zero modes of the operators $D_1$ and
$D_1^{\dag}$ are related to the zero modes of the operators
$D_1D_1^{\dag}$ and $D_1^{\dag}D_1$. Additionally, the zero modes
of the operators $D_1D_1^{\dag}$ and $D_1^{\dag}D_1$ can be
classified to parity positive and parity negative solutions
according to their Witten parity.

\noindent Note that the SUSY QM structure exists if
$\psi_{\bar{2}}=0$ on this matter curve. Moreover, SUSY is
unbroken, since $I_{D_{1}}\neq 0$ (the operator $D_1^{\dag}$ has
no localized zero modes).

\subsection{Localized fermion around $z_2=0$}

Along the curve $\Sigma_2$, we have $q_1=0$ and the fermions are
peaked around $z_2=0$. The localized solutions to the equations of
motion (\ref{eqmotion2}) read:
\begin{equation}\label{locali2}
\psi_{\bar{2}}=-\chi,{\,}{\,}{\,}{\,}{\,}{\,}\chi_2=g(z_2)e^{-q_2m^2|z_1|^2},{\,}{\,}{\,}\psi_{\bar{1}}=0.
\end{equation}
with $g(z_1)$ an arbitrary holomorphic function of $z_1$. The
$N=2$ SUSY QM algebra can be defined in terms of the $D_2$ matrix,
which is equal to:
\begin{equation}\label{dmatrix2344}
D_2=\left(%
\begin{array}{cc}
  \partial_2 & -m^2\bar{z}_2q_2 \\
  -m^2z_2q_2 & \bar{\partial}_{2} \\
\end{array}%
\right)
\end{equation}
acting on
\begin{equation}\label{sdfr}
\left(%
\begin{array}{c}
  \psi_{\bar{2}} \\
  \chi_2 \\
\end{array}%
\right)
\end{equation}

\subsection{Localized fermion around $z_1+z_2=0$}

The matter curve $\Sigma_3$, corresponds to generic charges $q_1$
and $q_2$. Performing the transformations:
\begin{align}\label{transforma}
&w=z_1+z_2,{\,}{\,}{\,}{\,}{\,}{\,}{\,}{\,}{\,}{\,}{\,}{\,}\psi_{\bar{w}}=\frac{1}{2}\large{(}\psi_{\bar{1}}+\psi_{\bar{2}}\large{)}
\\ & \notag u=z_1-z_2,{\,}{\,}{\,}{\,}{\,}{\,}{\,}{\,}{\,}{\,}{\,}{\,}\psi_{\bar{u}}=\frac{1}{2}\large{(}\psi_{\bar{1}}-\psi_{\bar{2}}\large{)}
\end{align}
the equations of motion (\ref{eqmotion2}) can be written:
\begin{align}\label{eqmotion3}
& 2{\,}\partial_w\psi_{\bar{w}}+2{\,}\partial_u\psi_{\bar{u}}
-\frac{m^2}{2}\big{(}\bar{w}(q_1+q_2)+\bar{u}(q_1-q_2)\big{)}\chi =0
\\  \notag & {\,}{\,}{\,}{\,}{\,}{\,}{\,}{\,}{\,}{\,}{\,}{\,}{\,}{\,}{\,}{\,}{\,}{\,}{\,}{\,}{\,}{\,}{\,}{\,}{\,}2{\,}\bar{\partial}_{\bar{w}}\chi-m^2\Big{(}w(q_1+q_2)+u(q_1-q_2)\Big{)}\psi_{\bar{w}} =0
\\  \notag & {\,}{\,}{\,}{\,}{\,}{\,}{\,}{\,}{\,}{\,}{\,}{\,}{\,}{\,}{\,}{\,}{\,}{\,}{\,}{\,}{\,}{\,}{\,}{\,}{\,}{\,}2{\,}\bar{\partial}_{\bar{u}}\chi-m^2\Big{(}w(q_1+q_2)+u(q_1-q_2)\Big{)}\psi_{\bar{u}} =0
\end{align}
When $\psi_{\bar{u}}=0$, an $N=2$ SUSY QM algebra underlies the
fermion system, defined in terms of the matrices $D_3$ and
$D^{\dag}_3$ as:
\begin{equation}\label{dmat}
D_3=\left(%
\begin{array}{cc}
  2{\,}\partial_w & -\frac{m^2}{2}\Big{(}\bar{w}(q_1+q_2)+\bar{u}(q_1-q_2)\Big{)} \\
  -m^2\Big{(}w(q_1+q_2)+u(q_1-q_2)\Big{)} & 2{\,}\bar{\partial}_w \\
\end{array}%
\right)
\end{equation}
and,
\begin{equation}\label{dmatr}
D^{\dag}_3=\left(%
\begin{array}{cc}
  2{\,}\bar{\partial}_w  & -m^2\Big{(}\bar{w}(q_1+q_2)+\bar{u}(q_1-q_2)\Big{)} \\
  -\frac{m^2}{2}\Big{(}w(q_1+q_2)+u(q_1-q_2)\Big{)} & 2{\,}\partial_w \\
\end{array}%
\right)
\end{equation}
acting on,
\begin{equation}\label{we23e}
\left(%
\begin{array}{c}
  \psi_{\bar{w}} \\
  \chi_w \\
\end{array}%
\right)
\end{equation}
Then, the fermionic localized solutions to the new equations of
motion (\ref{eqmotion3}) around $z_1+z_2=0$ are:
\begin{equation}\label{locali3}
\psi_{\bar{w}}=\frac{1}{\sqrt{2}}\chi,{\,}{\,}{\,}{\,}{\,}{\,}\chi_w=g(u)e^{-\frac{q_2m^2}{\sqrt{2}}|w|^2},{\,}{\,}{\,}\psi_{\bar{u}}=0.
\end{equation}
We  therefore conclude that each matter curve corresponds to an
underlying $N=2$ SUSY QM algebra. In turn, each SUSY algebra can
be constructed using the operators $D_1$, $D_2$ and $D_3$
respectively, the zero modes of which correspond to the solutions
of (\ref{eqmotion2}).

\subsection{Gravitational Backreaction on the Base Manifold-Metric Perturbations}

In the previous section we chose the canonical form for the metric
that describes $S$. However the surface $S$ is more like a base
space of the Calabi-Yau threefold and not a divisor \cite{f13}.
Therefore there is no way to know what metric describes precisely
the base space $S$, hence there is some freedom in the choice of
the metric on $S$. The metric adopted in the previous section is
the simplest case and describes perfectly the case for which the
system is fully described by an Super Yang-Mills theory, and
gravity is decoupled, as we previously noted. However we are free
to choose another metric that incorporates the gravitational
backreaction of the surface $S$ on the system. Note that the
volume of $S$ gives the gauge coupling of the effective
four-dimensional GUT \cite{eranpalti}. In this section we shall
put the previous section's index problem, into a different
context, by perturbing the metric of the complex surface $S$ in
the following way:
\begin{equation}\label{metric}
\mathrm{d}s^2=\big{(}1+\epsilon f_1(z_1)\big{)}
\mathrm{d}z_1\otimes\mathrm{d}\bar{z}_1+\big{(} 1+\epsilon
f_2(z_2)\big{)}\mathrm{d}z_2\otimes\mathrm{d}\bar{z}_2
\end{equation}
Using the above metric, the K\"{a}hler form is written as follows,
\begin{equation}\label{kahler1}
\omega =\frac{i}{2}\big{(}1+\epsilon f_1(z_1)\big{)}
\mathrm{d}z_1\wedge\mathrm{d}\bar{z}_1+\frac{i}{2}\big{(}1+\epsilon
f_2(z_2)\big{)} \mathrm{d}z_2\wedge\mathrm{d}\bar{z}_2
\end{equation}
The corresponding equations of motion for the fermionic fields
are:
\begin{align}\label{eqmotperturbed}
& \big{(}1+\epsilon
f_1(z_1)\big{)}\partial_2\psi_{\bar{2}}+\big{(}1+\epsilon
f_2(z_2)\big{)}\partial_1\psi_{\bar{1}}
-m^2(\bar{z}_1q_1+\bar{z}_2q_2)\chi =0
\\  \notag & {\,}{\,}{\,}{\,}{\,}{\,}{\,}{\,}{\,}{\,}{\,}{\,}{\,}{\,}{\,}{\,}{\,}{\,}{\,}{\,}{\,}{\,}{\,}{\,}{\,}{\,}{\,}{\,}{\,}{\,}{\,}{\,}{\,}{\,}{\,}{\,}{\,}{\,}{\,}{\,}{\,}{\,}{\,}{\,}{\,}{\,}{\,}{\,}{\,}{\,}{\,}{\,}{\,}{\,}{\,}{\,}{\,}{\,}{\,}{\,}{\,}{\,}{\,}{\,}{\,}{\,}{\,}{\,}{\,}{\,}{\,}{\,}{\,}{\,}{\,}{\,}{\,}{\,}{\,}{\,}{\,}{\,}{\,}{\,}{\,}{\,}{\,}{\,}\bar{\partial}_1\chi-m^2(z_1q_1+z_2q_2)\psi_{\bar{1}} =0
\\  \notag & {\,}{\,}{\,}{\,}{\,}{\,}{\,}{\,}{\,}{\,}{\,}{\,}{\,}{\,}{\,}{\,}{\,}{\,}{\,}{\,}{\,}{\,}{\,}{\,}{\,}{\,}{\,}{\,}{\,}{\,}{\,}{\,}{\,}{\,}{\,}{\,}{\,}{\,}{\,}{\,}{\,}{\,}{\,}{\,}{\,}{\,}{\,}{\,}{\,}{\,}{\,}{\,}{\,}{\,}{\,}{\,}{\,}{\,}{\,}{\,}{\,}{\,}{\,}{\,}{\,}{\,}{\,}{\,}{\,}{\,}{\,}{\,}{\,}{\,}{\,}{\,}{\,}{\,}{\,}{\,}{\,}{\,}{\,}{\,}{\,}{\,}{\,}{\,}\bar{\partial}_2\chi-m^2(z_1q_1+z_2q_2)\psi_{\bar{2}} =0
\end{align}
By looking at the equations of motion (\ref{eqmotperturbed}), we
can generally say that the form of the localized solutions along
each matter curve will have a more evolved dependence on all the
local coordinates that parameterize the complex surface $S$. By
looking equation (\ref{metric}) we can see that the functions
$f_1,f_2$ have a holomorphic dependence on their coordinates.
There is a particular reason for using holomorphic functions,
which is the fact that the solutions of the equations of motions
(wave functions) are the sections of holomorphic line bundles
along the loci $z_1=0$, $z_2=0$ and $z_1+z_2=0$ \cite{f13}. In
this section we shall study if the holomorphic linear perturbation
of the metric (\ref{metric}) modifies the spectral problem of the
operator corresponding to each matter curve. However we shall not
be interested in the particular form of the localized wave
functions that solve the equations of motion. Additionally, due to
the lack of knowledge of the global geometry that describes the
compact complex threefold (also since the local geometry around
the singularity affects the Standard Model physics), and in order
to avoid theoretical inconsistencies, we assume that the functions
$f_1$ and $f_2$ are decreasing functions of their arguments.

\subsubsection{The matter curve $z_1=0$}

Let us start with the matter curve $z_1=0$, which means that
$q_2=0$. By using the holomorphic perturbation of the metric
(\ref{metric}) we can see that, the whole problem is a
perturbation of the one that corresponds to the canonical metric.
Indeed, as can be easily checked, localized solutions can exist if
$\psi_{\bar{2}}=0$ (the situation is similar to the un-perturbed
case). Then, by setting $q_2=0$, the equations of motion
corresponding to the matter curve $z_1=0$, are:
\begin{align}\label{eqmotperturbed1}
& \big{(}1+\epsilon
f_2(z_2,\bar{z}_2)\big{)}\partial_1\psi_{\bar{1}}
-m^2\bar{z}_1q_1\chi =0
\\  \notag & {\,}{\,}{\,}{\,}{\,}{\,}{\,}{\,}{\,}{\,}{\,}{\,}{\,}{\,}{\,}{\,}{\,}{\,}{\,}{\,}{\,}{\,}{\,}{\,}{\,}{\,}{\,}{\,}{\,}{\,}{\,}{\,}{\,}{\,}{\,}{\,}{\,}{\,}{\,}{\,}{\,}{\,}\bar{\partial}_1\chi-m^2z_1q_1\psi_{\bar{1}} =0
\end{align}
which can be recast as,
\begin{align}\label{eqmotperturbed12}
& \partial_1\psi_{\bar{1}}
-\frac{m^2\bar{z}_1q_1}{\big{(}1+\epsilon
f_2(z_2,\bar{z}_2)\big{)}}\chi =0
\\  \notag & {\,}{\,}{\,}{\,}{\,}{\,}{\,}{\,}{\,}{\,}{\,}{\,}{\,}{\,}{\,}{\,}{\,}{\,}{\,}{\,}{\,}{\,}{\,}{\,}{\,}\bar{\partial}_1\chi-m^2z_1q_1\psi_{\bar{1}} =0
\end{align}
Performing a perturbation expansion and keeping terms linear to
the expansion parameter $\epsilon$ we obtain:
\begin{align}\label{eqmotperturbed14}
& \partial_1\psi_{\bar{1}} -m^2\bar{z}_1q_1\big{(}1-\epsilon
f_2(z_2,\bar{z}_2)\big{)}\chi =0
\\  \notag & {\,}{\,}{\,}{\,}{\,}{\,}{\,}{\,}{\,}{\,}{\,}{\,}{\,}{\,}{\,}{\,}{\,}{\,}{\,}{\,}{\,}{\,}{\,}{\,}{\,}{\,}{\,}{\,}{\,}{\,}{\,}{\,}{\,}{\,}{\,}{\,}{\,}{\,}{\,}{\,}{\,}{\,}\bar{\partial}_1\chi-m^2z_1q_1\psi_{\bar{1}} =0
\end{align}
Clearly, the zero modes of the above equation
(\ref{eqmotperturbed14}) correspond to the zero modes of the
matrix:
\begin{equation}\label{matrixodd}
D_{1\epsilon}=\left(%
\begin{array}{cc}
  \partial_1 & -m^2\bar{z}_1q_1\big{(}1-\epsilon
f_2(z_2,\bar{z}_2)\big{)} \\
  -m^2z_1q_1 & \bar{\partial}_{1} \\
\end{array}%
\right)
\end{equation}
We can write $D_{1\epsilon}=D_1+C$, with $D_1$ as in equation
(\ref{dmatrix1}) and $C$ being the matrix:
\begin{equation}\label{codd}
C=\left(%
\begin{array}{cc}
  0 & m^2\bar{z}_1q_1\epsilon
f_2(z_2,\bar{z}_2) \\
  0 & 0 \\
\end{array}%
\right)
\end{equation}
There exists a theorem in the mathematical literature that
guarantees invariance of the index of Fredholm operators under odd
perturbations of Fredholm type
\cite{thaller,theoremodd,oikonomou}. Particularly the theorem
states:
\begin{itemize}
\item[*]Let $Q$ be a Fredholm operator and $C$ be an odd operator.
Then, $Q+C$ is a Fredholm operator then the indices of the two
operators are equal, i.e. :
\begin{equation}\label{indexfredtheorodd}
\mathrm{ind}(D_{1}+C)=\mathrm{ind}D_{1\epsilon}
\end{equation}
\end{itemize}
We must note that an odd operator is defined as a matrix that
anti-commutes with the Witten operator, $W$, that is $\{W,C\}=0$.
Using the notation we introduced in section 3, the matrix $W$ is
equal to:
\begin{equation}\label{smatrixwad}
W=\bigg{(}\begin{array}{ccc}
  1 & 0 \\
  0 & -1  \\
\end{array}\bigg{)}
\end{equation}
It can be easily seen that the matrix $C$, defined in equation
(\ref{codd}) is odd (using the terminology of the theorem), since
it anti-commutes with $W$. Therefore the indices of the two
matrices $D_{1}+C$ and $D_{1}$ are equal, that is,
\begin{equation}\label{indexfredtheorodrgtd}
\mathrm{ind}(D_{1\epsilon}+C)=\mathrm{ind}D_{1\epsilon}
\end{equation}
As a consequence of the aforementioned results, the Witten index
of the composite operator $D_{1}+C$  is equal to the Witten index
of the operator $D_{1}$. A direct implication of the equality of
the two indices is that the spectral problem of the two operators
is the same. This does not necessarily imply that the zero modes
of $D_{1}$ is equal to the zero modes of the operator $D_{1}+C$,
but it certainly implies that the net number of the zero modes
corresponding to the operators and their adjoint are equal. This
is of particular importance since it gives us the opportunity to
study more evolved cases and investigate more difficult aspects of
these problems, such as the spectral asymmetry of the operators.

\noindent The above result does not change if we include higher orders of $\epsilon$ in the matrix $C$. Indeed, the matrix $C$ would then be:
\begin{equation}\label{matrixchigher}
C=\left(%
\begin{array}{cc}
  0 & m^2\bar{z}_1q_1\epsilon
f_2(z_2,\bar{z}_2)- m^2\bar{z}_1q_1\epsilon^2
f_2^2(z_2,\bar{z}_2)+... \\
  0 & 0 \\
\end{array}%
\right)
\end{equation}
which still satisfies the theorem above.

\noindent Note that the situation we studied in this section can be much more difficult in
the case a background flux is turned on. In that case, the restrictions on K\"{a}hler form are
more stringent, since the K\"{a}hler form must satisfy the D-term equation:
\begin{equation}\label{Dtermeqn}
i[\phi,\bar{\phi}]+2\omega \wedge F^{1,1}+*sD=0
\end{equation}
where in the above $F^{1,1}$ stands for the flux.

\subsubsection{The matter curve $z_2=0$}

In the case of the $z_2=0$ matter curve, we have $q_2=0$. As a
result of the holomorphicity of the function $f_1(z_1)$, in order
to solve the equations of motion, we must set $\psi_{\bar{1}}=0$
just in the non-perturbed case. Then, the equations of motion are
written,
\begin{align}\label{eqmotperturbed14}
& \partial_2\psi_{\bar{2}} -m^2\bar{z}_2q_2\big{(}1-\epsilon
f_2(z_2,\bar{z}_2)\big{)}\chi =0
\\  \notag & {\,}{\,}{\,}{\,}{\,}{\,}{\,}{\,}{\,}{\,}{\,}{\,}{\,}{\,}{\,}{\,}{\,}{\,}{\,}{\,}{\,}{\,}{\,}{\,}{\,}{\,}{\,}{\,}{\,}{\,}{\,}{\,}{\,}{\,}{\,}{\,}{\,}{\,}{\,}{\,}{\,}{\,}\bar{\partial}_2\chi-m^2z_2q_2\psi_{\bar{2}} =0
\end{align}
As in the $z_1=0$ case, we can write $D_{2\epsilon}=D_2+C$, with
$D_{2\epsilon}$ being,
\begin{equation}\label{matrixodd}
D_{2\epsilon}=\left(%
\begin{array}{cc}
  \partial_2 & -m^2\bar{z}_2q_2\big{(}1-\epsilon
f_1(z_1,\bar{z}_1)\big{)} \\
  -m^2z_2q_2 & \bar{\partial}_{2} \\
\end{array}%
\right)
\end{equation}
and $D_2$ as in equation (\ref{dmatrix2344}). In this case the
matrix $C$ is equal to:
\begin{equation}\label{codd}
C=\left(%
\begin{array}{cc}
  0 & m^2\bar{z}_2q_2\epsilon
f_1(z_1,\bar{z}_1) \\
  0 & 0 \\
\end{array}%
\right)
\end{equation}
Both the matrices $C$ and $D_2$ satisfy the requirements of the
theorem we used previously, therefore we also have in this case:
\begin{equation}\label{indexfredtheoezel}
\mathrm{ind}(D_{2}+C)=\mathrm{ind}D_{2\epsilon}
\end{equation}

\subsubsection{The Higgs curve $z_1+z_2=0$}

The case $z_1+z_2=0$ is much more evolved than the previous two
cases. Using the transformations (\ref{transforma}), equation
(\ref{eqmotperturbed}) can be cast as:
\begin{align}\label{eqmotion4perturbed}
& (2+\epsilon f_1+\epsilon
f_2){\,}\partial_w\psi_{\bar{w}}+(2+\epsilon f_1+\epsilon
f_2){\,}\partial_u\psi_{\bar{u}}
\\ \notag &+(\epsilon f_2-\epsilon f_1)(\partial_w\psi_u+\partial_u\psi_w)
-\frac{m^2}{2}\big{(}\bar{w}(q_1+q_2)+\bar{u}(q_1-q_2)\big{)}\chi
=0
\\  \notag & {\,}{\,}{\,}{\,}{\,}{\,}{\,}{\,}{\,}{\,}{\,}{\,}{\,}{\,}{\,}{\,}{\,}{\,}{\,}{\,}{\,}{\,}{\,}{\,}{\,}{\,}{\,}{\,}{\,}{\,}{\,}{\,}{\,}{\,}{\,}{\,}{\,}{\,}{\,}{\,}{\,}{\,}{\,}{\,}{\,}{\,}{\,}{\,}{\,}{\,}{\,}{\,}{\,}{\,}{\,}{\,}{\,}2{\,}\bar{\partial}_{\bar{w}}\chi-m^2\Big{(}w(q_1+q_2)+u(q_1-q_2)\Big{)}\psi_{\bar{w}} =0
\\  \notag & {\,}{\,}{\,}{\,}{\,}{\,}{\,}{\,}{\,}{\,}{\,}{\,}{\,}{\,}{\,}{\,}{\,}{\,}{\,}{\,}{\,}{\,}{\,}{\,}{\,}{\,}{\,}{\,}{\,}{\,}{\,}{\,}{\,}{\,}{\,}{\,}{\,}{\,}{\,}{\,}{\,}{\,}{\,}{\,}{\,}{\,}{\,}{\,}{\,}{\,}{\,}{\,}{\,}{\,}{\,}{\,}{\,}{\,}{\,}2{\,}\bar{\partial}_{\bar{u}}\chi-m^2\Big{(}w(q_1+q_2)+u(q_1-q_2)\Big{)}\psi_{\bar{u}} =0
\end{align}
In this case the theorem we presented previously does not find
application, since the complex derivatives are interrelated. The
only case that the theorem can find application is when $f_1=f_2$.
Nevertheless, the last case corresponds to a trivial (coordinate
independent) deformation of the metric, thus it is a perturbative
constant shift. In the same way as in the un-perturbed $z_1+z_2=0$
case, when $\psi_{\bar{u}}=0$ and $f_1=f_2=f$, the above equation
can be cast as:
\begin{align}\label{eqmotion4perturbed1}
& (2+2\epsilon f){\,}\partial_w\psi_{\bar{w}}
-\frac{m^2}{2}\big{(}\bar{w}(q_1+q_2)+\bar{u}(q_1-q_2)\big{)}\chi =0
\\  \notag & {\,}{\,}{\,}{\,}{\,}{\,}{\,}{\,}{\,}{\,}{\,}{\,}{\,}{\,}{\,}{\,}{\,}{\,}{\,}{\,}{\,}{\,}{\,}{\,}{\,}2{\,}\bar{\partial}_{\bar{w}}\chi-m^2\Big{(}w(q_1+q_2)+u(q_1-q_2)\Big{)}\psi_{\bar{w}} =0
\end{align}
Following the same steps as previously, we obtain:
\begin{align}\label{eqmotion4perturbed11}
& 2{\,}\partial_w\psi_{\bar{w}}
-(1-\epsilon f)\frac{m^2}{2}\big{(}\bar{w}(q_1+q_2)+\bar{u}(q_1-q_2)\big{)}\chi =0
\\  \notag & {\,}{\,}{\,}{\,}{\,}{\,}{\,}{\,}{\,}{\,}{\,}{\,}{\,}{\,}{\,}{\,}{\,}{\,}{\,}{\,}{\,}{\,}{\,}{\,}{\,}2{\,}\bar{\partial}_{\bar{w}}\chi-m^2\Big{(}w(q_1+q_2)+u(q_1-q_2)\Big{)}\psi_{\bar{w}} =0
\end{align}
The zero modes of the above equation are the zero modes of the
matrix:
\begin{equation}\label{matrixodd}
D_{w\epsilon}=\left(%
\begin{array}{cc}
 2{\,}\partial_w & -\frac{m^2}{2}\big{(}\bar{w}(q_1+q_2)+\bar{u}(q_1-q_2)\big{)}(1-\epsilon f) \\
  -m^2z_1q_1 & \bar{\partial}_{1} \\
\end{array}%
\right)
\end{equation}
Likewise, we can write $D_{w\epsilon}=D_w+C_w$, with $D_w$ as in
equation (\ref{dmat}) and $C_w$:
\begin{equation}\label{codd}
C_w=\left(%
\begin{array}{cc}
  0 & \frac{m^2}{2}\big{(}\bar{w}(q_1+q_2)+\bar{u}(q_1-q_2)\big{)}\epsilon f \\
  0 & 0 \\
\end{array}%
\right)
\end{equation}
Therefore applying the theorem for the two matrices, we have:
\begin{equation}\label{indexfredtheorodd}
\mathrm{ind}(D_{w}+C)=\mathrm{ind}D_{w\epsilon}
\end{equation}
Hence, the indices of the two operators are equal.

The results of this section are very important since, in virtue of
the theorem, the net number of the zero modes of the
metric-perturbed fermionic system is equal to the net number of
the zero modes that the Euclidean metric-fermionic system has.
Nevertheless we know that the solutions exist, but this theorem
tells us nothing on how these perturbed solutions behave. Before
we close this section, we must note that in the case we perform a
non-holomorphic perturbation of the Euclidean metric, the
solutions of the equation of motion are not the ones that appeared
in this section. Indeed, let us take for example the matter curve
$z_1=0$, for which a non holomorphic perturbation of the metric
would result to three wave functions-solutions to the equation of
motion, namely $\chi$, $\psi_{\bar{1}}$ and $\psi_{\bar{2}}$. The
solutions $\psi_{\bar{1}}$ and $\psi_{\bar{2}}$ are given as
functions of $\chi$, which in turn is a perturbation of the
gaussian profile solution. For a specific example of this type,
see for example reference \cite{eranpalti}.

\section{Yukawa Couplings in the Presence of Constant Background Gauge Fluxes and SUSY QM}

The situation of the fermionic system without background gauge
fluxes is very useful but we can get only one non-trivial Yukawa
coupling \cite{f1}. In order to obtain the hierarchies of the
quark masses and the appropriate mixing of the quark and lepton
matter fields, the wave functions we found in section 3 must be
appropriately distorted \cite{f1}. This distortion can be caused
by the appearance of background gauge fields. It is proven that
when the gauge fluxes are field dependent, then reasonable
agreement with the observed mass hierarchies and mixings can be
achieved \cite{f1}. In this section we shall add non-trivial
background gauge fluxes and study whether the resulting localized
fields on each matter curve on $S$ are related to an $N=2$ SUSY QM
algebra. We shall follow reference \cite{f1}. Trying to find
localized solutions along the matter curve, when the gauge fields
have a local coordinate dependence can be quite difficult. We
shall confine ourselves to the case where the gauge fields are
constant and independent from the coordinates $z_1,z_2$.

In general, the total flux can be written as follows \cite{f1}:
\begin{equation}\label{totalflux}
\mathcal{F}=FQ+F^{(1)}Q_1+F^{(2)}Q_2
\end{equation}
In the above equation, $\mathcal{F}$ is the total flux, $F$ is the
$U(1)$ bulk gauge flux, with generator $Q$, and $F^{(1)}$,
$F^{(2)}$ are the fluxes along the matter curves $z_1$ and $z_2$
respectively (with generators $Q_1$ and $Q_2$ as we saw in section
3). The corresponding gauge potentials are $\mathcal{A}$, $A$ and
$A^{(1)}$, $A^{(2)}$, respectively, with,
\begin{equation}\label{totalgaugepotential}
\mathcal{A}=qA+q_1A^{(1)}+q_2A^{(2)}
\end{equation}
In the above, $q$ stands for the total $U(1)$ bulk charge, $q_1$
is the $U(1)$ charge along the matter curve $z_1$ and the and
$q_2$ is the $U(1)$ charge along the matter curve $z_2$. The bulk
flux breaks the initial $G_s$ gauge symmetry to $\Gamma_s\times
U(1)$, and the fermions transform to a representation $R$ which a
direct sum of irreducible representations labelled as
$(q,q_1,q_2)$. In the general case, and if we consider only
diagonal components of the gauge flux, the bulk flux can be
written \cite{f1}:
\begin{equation}\label{nonconsgaugeneurol}
F=F_{1\bar{1}}\mathrm{d}z_1\wedge
\mathrm{d}\bar{z}_1+F_{2\bar{2}}\mathrm{d}z_2\wedge
\mathrm{d}\bar{z}_2
\end{equation}
and the $U(1)$'s along the matter curves are taken to be:
\begin{equation}\label{nonconfluxesalongmattercurves}
F^{(1)}=F_{2\bar{2}}^{(1)}\mathrm{d}z_2\wedge
\mathrm{d}\bar{z}_2,{\,}{\,}{\,}{\,}{\,}{\,}{\,}{\,}F^{(2)}=F_{1\bar{1}}^{(2)}\mathrm{d}z_1\wedge
\mathrm{d}\bar{z}_1
\end{equation}
Hence, if the adjoint vacuum expectation value $\langle
\phi\rangle$ is the same as in equation (\ref{backscalar1}), the
equations of motion for the charged fermionic fields are
\cite{f1}:
\begin{align}\label{eqmgaugecase11111}
&
(\partial_2-iA_2){\psi}_{\bar{2}}+(\partial_1-iA_1){\psi}_{\bar{1}}
-m^2(\bar{z}_1q_1+\bar{z}_2q_2){\chi} =0
\\  \notag & {\,}{\,}{\,}{\,}{\,}{\,}{\,}{\,}{\,}{\,}{\,}{\,}{\,}{\,}{\,}{\,}{\,}{\,}{\,}(\partial_2-iA_{\bar{1}})\chi-m^2(z_1q_1+z_2q_2){\psi}_{\bar{1}} =0
\\  \notag & {\,}{\,}{\,}{\,}{\,}{\,}{\,}{\,}{\,}{\,}{\,}{\,}{\,}{\,}{\,}{\,}{\,}{\,}{\,}(\partial_2-iA_{\bar{2}}){\chi}-m^2(z_1q_1+z_2q_2){\psi}_{\bar{2}} =0
\end{align}
In the constant gauge flux case, we take:
\begin{equation}\label{consgaugeneurol}
F=2iM\mathrm{d}z_1\wedge
\mathrm{d}\bar{z}_1+2iN\mathrm{d}z_2\wedge \mathrm{d}\bar{z}_2
\end{equation}
with $M$, $N$, real constants. The fluxes along the matter curves
are then equal to:
\begin{equation}\label{fluxesalongmattercurves}
F^{(1)}=2iN^{(1)}\mathrm{d}z_2\wedge
\mathrm{d}\bar{z}_2,{\,}{\,}{\,}{\,}{\,}{\,}{\,}{\,}F^{(2)}=2iM^{(2)}\mathrm{d}z_1\wedge
\mathrm{d}\bar{z}_1
\end{equation}
where $N^{(1)}$ and $M^{(2)}$ real constants. Therefore, the gauge
potentials are equal to:
\begin{align}\label{gpwerr}
&
A=iM(z_1\mathrm{d}\bar{z}_1-\bar{z}_1\mathrm{d}z_1)+iN(z_2\mathrm{d}\bar{z}_2-\bar{z}_2\mathrm{d}z_2)
\\ \notag & A^{(1)}=iN^{(1)}(z_2\mathrm{d}\bar{z}_2-\bar{z}_2\mathrm{d}z_2)
\\ \notag & A^{(2)}=iM^{(2)}(z_1\mathrm{d}\bar{z}_1-\bar{z}_1\mathrm{d}z_1)
\end{align}
Consequently, the total gauge potential is equal to:
\begin{equation}\label{totalgaugepot}
\mathcal{A}=i(qM+q_2M^{(2)})(z_1\mathrm{d}\bar{z}_1-\bar{z}_1\mathrm{d}z_1)+i(qN+q_1N^{(1)})(z_2\mathrm{d}\bar{z}_2-\bar{z}_2\mathrm{d}z_2)
\end{equation}
Performing a suitable gauge transformation of the form,
\begin{equation}\label{gaugetransformation}
\mathcal{A}=\widehat{\mathcal{A}}+\mathrm{d}\Omega
\end{equation}
we can set $A_{\bar{1}}$=0 and $A_{\bar{2}}$=0 in equation
(\ref{eqmgaugecase11111}) and work with the hatted fields. Indeed,
equation (\ref{eqmgaugecase11111}) can simplified to:
\begin{align}\label{eqmgaugecase}
&
(\partial_2-i\widehat{A}_2)\widehat{\psi}_{\bar{2}}+(\partial_1-i\widehat{A}_1)\widehat{\psi}_{\bar{1}}
-m^2(\bar{z}_1q_1+\bar{z}_2q_2)\widehat{\chi} =0
\\  \notag & {\,}{\,}{\,}{\,}{\,}{\,}{\,}{\,}{\,}{\,}{\,}{\,}{\,}{\,}{\,}{\,}{\,}{\,}{\,}\bar{\partial}_1\widehat{\chi}-m^2(z_1q_1+z_2q_2)\widehat{\psi}_{\bar{1}} =0
\\  \notag & {\,}{\,}{\,}{\,}{\,}{\,}{\,}{\,}{\,}{\,}{\,}{\,}{\,}{\,}{\,}{\,}{\,}{\,}{\,}\bar{\partial}_2\widehat{\chi}-m^2(z_1q_1+z_2q_2)\widehat{\psi}_{\bar{2}} =0
\end{align}
with $\chi=e^{i\Omega}\widehat{\chi}$,
$\psi_{\bar{1}}=e^{i\Omega}\widehat{\psi}_{\bar{1}}$ and
$\psi_{\bar{2}}=e^{i\Omega}\widehat{\psi}_{\bar{2}}$. Supposing
that the gauge field is coordinate independent, the total gauge
potential reads:
\begin{equation}\label{constantflux}
\widehat{\mathcal{A}}=-2iM\bar{z}_1\mathrm{d}z_1-2iN\bar{z}_2\mathrm{d}z_2
\end{equation}
and $\widehat{A}_1=-2iM\bar{z}_1$ and
$\widehat{A}_2=-2iN\bar{z}_2$. The gauge parameter $\Omega$ is in
this case:
\begin{equation}\label{omegaparame}
\Omega =i(M|z_1|^2+N|z_2|^2)
\end{equation}
Working in the gauge we chose above makes the calculation of the
wave functions (and hence of the corresponding gauge invariant
properties such as Yukawa couplings) simpler \cite{f1}. This gauge
is referred to as holomorphic gauge \cite{f1}.

\subsection{The Matter Curve $z_1=0$}

Let us study here the first matter curve $z_1=0$. By taking
$q_2=0$, the localized solutions in this case are \cite{f1}:
\begin{equation}\label{gaugesolutions1}
\widehat{\psi}_{\bar{1}}=-\frac{\lambda_1}{q_1m^2}\widehat{\chi},{\,}{\,}{\,}{\,}{\,}{\,}\widehat{\chi}=g(z_2)e^{-\lambda_1m^2|z_1|^2},{\,}{\,}{\,}\widehat{\psi}_{\bar{2}}=0.
\end{equation}
with $\lambda_1$ equal to:
\begin{equation}\label{lambda1}
\lambda_1=-M+q_1m^2\sqrt{1+\frac{M^2}{q_1^2m^4}}
\end{equation}
The above solutions correspond to the following equations of
motion:
\begin{align}\label{eqmgaugecase}
& (\partial_1-i\widehat{A}_1)\widehat{\psi}_{\bar{1}}
-m^2\bar{z}_1q_1\widehat{\chi} =0
\\  \notag & {\,}{\,}{\,}{\,}{\,}{\,}{\,}{\,}{\,}{\,}{\,}{\,}{\,}{\,}{\,}{\,}{\,}{\,}{\,}\bar{\partial}_1\widehat{\chi}-m^2z_1q_1\widehat{\psi}_{\bar{1}} =0
\end{align}
It is very easy to prove that we can associate an $N=2$ SUSY QM algebra corresponding to the equations of motion (\ref{eqmgaugecase}). Indeed,
following the steps of section 3 we define the matrix $D_{A_1}$ and also $D^{\dag}_{A_1}$ as follows,
\begin{equation}\label{dmatr655656ix1}
D_{A_1}=\left(%
\begin{array}{cc}
  \partial_1-i\widehat{A}_1 & -m^2\bar{z}_1q_1 \\
  -m^2z_1q_1 & \bar{\partial}_{1} \\
\end{array}%
\right)
\end{equation}
and,
\begin{equation}\label{dmat54465rix23}
D^{\dag}_{A_1}=\left(%
\begin{array}{cc}
  \bar{\partial}_1+i\widehat{\bar{A}}_1 & -m^2z_1q_1 \\
  -m^2\bar{z}_1q_1 & \partial_{1} \\
\end{array}%
\right)
\end{equation}
acting on,
\begin{equation}\label{we43435e33}
\left(%
\begin{array}{c}
  \widehat{\psi}_{\bar{1}} \\
  \widehat{\chi} \\
\end{array}%
\right)
\end{equation}
In the above $\widehat{A}_1=-2iM\bar{z}_1$. The matrix
$D^{\dag}_{A_1}$ has no zero modes, while the matrix $D_{A_1}$ has
solutions the functions of equation (\ref{gaugesolutions1}).
Therefore the Fredholm index $I_{D_A}$, of the operator $D_{A_1}$,
is equal to,
\begin{equation}\label{index34}
\mathrm{ind}\mathrm{I}_{D_A}=\mathrm{dim{\,}ker}(D^{\dag}_{A_1})-\mathrm{dim{\,}ker}(D_{A_1})
\end{equation}
for which clearly $\mathrm{ind}\mathrm{I}_{D_A}\neq 0$. From the
last we conclude that SUSY is unbroken.

\noindent The $N=2$ supersymmetric
quantum mechanical system can be defined using the supercharges $Q_A$ and
$Q^{\dag}_A$,
\begin{equation}\label{wit2134}
Q=\bigg{(}\begin{array}{ccc}
  0 & D_{A_1} \\
  0 & 0  \\
\end{array}\bigg{)},{\,}{\,}{\,}{\,}{\,}{\,}{\,}{\,}Q^{\dag}=\bigg{(}\begin{array}{ccc}
  0 & 0 \\
  D^{\dag}_{A_1} & 0  \\
\end{array}\bigg{)}
\end{equation}
Furthermore, the Hamiltonian can be written as:
\begin{equation}\label{wit4123}
H=\bigg{(}\begin{array}{ccc}
  D_{A_1}D_{A_1}^{\dag} & 0 \\
  0 & D_{A_1}^{\dag}D_{A_1}  \\
\end{array}\bigg{)}
\end{equation}
Finally, the Witten index of the SUSY QM algebra, is
$I_{D_{A}}=-\Delta$.

We can see that the constant background gauge fluxes do not spoil
the SUSY QM algebra that underlies the fermionic system of the
flux-less case. The algebra itself is of-course different but
still SUSY is unbroken.

\subsection{The Matter Curve $z_2=0$}

In the case of the $z_2=0$ matter curve, the equations of motion are (for $q_1=0$):

\begin{align}\label{eqmgaugecase1}
& (\partial_2-iA_2)\widehat{\psi}_{\bar{2}}
-m^2\bar{z}_2q_2\widehat{\chi} =0
\\  \notag & {\,}{\,}{\,}{\,}{\,}{\,}{\,}{\,}{\,}{\,}{\,}{\,}{\,}{\,}{\,}{\,}{\,}{\,}{\,}\bar{\partial}_2\widehat{\chi}-m^2z_2q_2\widehat{\psi}_{\bar{2}} =0
\end{align}
with $A_2=-2iN\bar{z}_2$. The localized solutions are:
\begin{equation}\label{gaugesolutions2}
\widehat{\psi}_{\bar{2}}=-\frac{\lambda_2}{q_2m^2}\widehat{\chi},{\,}{\,}{\,}{\,}{\,}{\,}\widehat{\chi}=g(z_1)e^{-\lambda_2m^2|z_2|^2},{\,}{\,}{\,}\widehat{\psi}_{\bar{1}}=0.
\end{equation}
with $\lambda_2$ equal to:
\begin{equation}\label{lambda1}
\lambda_2=-N+q_2m^2\sqrt{1+\frac{N^2}{q_2^2m^4}}
\end{equation}
The $N=2$ SUSY QM algebra is built on the matrices:
\begin{equation}
D_{A_{2}}=\left(%
\begin{array}{cc}
  \partial_2-iA_2 & -m^2\bar{z}_2q_2 \\
  -m^2z_2q_2 & \bar{\partial}_{2} \\
\end{array}%
\right)
\end{equation}
and,
\begin{equation}\label{dmat54465rix23}
D^{\dag}_{A_2}=\left(%
\begin{array}{cc}
  \bar{\partial}_2+i\bar{A}_2 & -m^2z_2q_2 \\
  -m^2\bar{z}_2q_2  & \partial_{2} \\
\end{array}%
\right)
\end{equation}
acting on,
\begin{equation}\label{we43435e33}
\left(%
\begin{array}{c}
  \widehat{\psi}_{\bar{2}} \\
  \widehat{\chi} \\
\end{array}%
\right)
\end{equation}
We shall not pursuit this case further, since it is identical with the previous $z_1=0$. The result is that a $N=2$ unbroken SUSY QM algebra underlies the system.

As for the $z_1+z_2=0$ case, it is much more difficult to handle,
compared to the other two cases. Performing the transformation
(\ref{transforma}), the equations of motion (\ref{eqmgaugecase})
can be cast in the form:
\begin{align}\label{eqmotion36ggfd}
&
2{\,}\partial_w\widehat{\psi}_{\bar{w}}+2{\,}\partial_u\widehat{\psi}_{\bar{u}}+\big{(}-N(\bar{w}-\bar{u})-M(\bar{w}+\bar{u})\big{)}\widehat{\psi}_{\bar{w}}
\\ \notag & +\big{(}N(\bar{w}-\bar{u})-M(\bar{w}+\bar{u})\big{)}\widehat{\psi}_{\bar{u}}
-\frac{m^2}{2}\big{(}\bar{w}(q_1+q_2)+\bar{u}(q_1-q_2)\big{)}\widehat{\chi}
=0
\\  \notag & {\,}{\,}{\,}{\,}{\,}{\,}{\,}{\,}{\,}{\,}{\,}{\,}{\,}{\,}{\,}{\,}{\,}{\,}{\,}{\,}{\,}{\,}{\,}{\,}{\,}{\,}{\,}{\,}{\,}{\,}{\,}{\,}{\,}{\,}{\,}{\,}{\,}{\,}{\,}{\,}{\,}{\,}{\,}{\,}{\,}{\,}{\,}{\,}{\,}{\,}{\,}{\,}{\,}{\,}{\,}{\,}{\,}{\,}{\,}{\,}{\,}2{\,}\bar{\partial}_{\bar{w}}\widehat{\chi}-m^2\Big {(}w(q_1+q_2)+u(q_1-q_2)\Big{)}\widehat{\psi}_{\bar{w}} =0
\\  \notag & {\,}{\,}{\,}{\,}{\,}{\,}{\,}{\,}{\,}{\,}{\,}{\,}{\,}{\,}{\,}{\,}{\,}{\,}{\,}{\,}{\,}{\,}{\,}{\,}{\,}{\,}{\,}{\,}{\,}{\,}{\,}{\,}{\,}{\,}{\,}{\,}{\,}{\,}{\,}{\,}{\,}{\,}{\,}{\,}{\,}{\,}{\,}{\,}{\,}{\,}{\,}{\,}{\,}{\,}{\,}{\,}{\,}{\,}{\,}{\,}{\,}{\,}2{\,}\bar{\partial}_{\bar{u}}\widehat{\chi}-m^2\Big{(}w(q_1+q_2)+u(q_1-q_2)\Big{)}\widehat{\psi}_{\bar{u}} =0
\end{align}
It is not easy to relate the above fermionic system to an $N=2$
SUSY QM algebra. Perhaps central charges must be included to this
$N=2$ algebra. Such a behavior kind of surprised us, because we
expected all the localized fermion solutions to have the same,
central charge free, $N=2$ SUSY QM algebra. It seems that this is
not the case. We shall not pursuit this issues further.

\section{Yukawa couplings in the absence of gauge fluxes and $N=2$ SUSY QM algebra.}

In the previous we found that when a matter curve has localized
zero modes, we can built a $N=2$ SUSY QM algebra from the system.
In most cases localization occurs when, one of the fields that
exist on the D7 brane intersection vanishes. By looking the
equations of motion (\ref{eqmotion3}), it is natural to make the
equations of motion look like the following,
\begin{align}\label{eqmotion43dgf}
&\big{(}2{\,}\partial_w+2{\,}\partial_u\big{)}\big{(}\psi_{\bar{w}}+\psi_{\bar{u}}\big{)}
-\frac{m^2}{2}\Big{(}\bar{w}(q_1+q_2)+\bar{u}(q_1-q_2)\Big{)}\chi
=0
\\  \notag & {\,}\big{(}2{\,}\bar{\partial}_{\bar{w}}+2{\,}\bar{\partial}_{\bar{u}}\big{)}\chi-m^2\Big{(}w(q_1+q_2)+u(q_1-q_2)\Big{)}\big{(}\psi_{\bar{w}}+\psi_{\bar{u}}\big{)} =0
\end{align}
This is clarified by looking the $\chi$ derivative, which is
$2\bar{\partial}_{\bar{w}}+2{\,}\bar{\partial}_{\bar{u}}$. In the
equations of motion (\ref{eqmotion43dgf}), the derivative that
acts on $\chi$ is the conjugate derivative of the one that acts on
$\psi_{\bar{w}}+\psi_{\bar{u}}$ \footnote{Recall the SUSY QM
algebras for the matter curves $z_1=0$ and $z_2=0$}. Let us see
when this is possible and what would be the implications of this
construction.

\noindent An $N=2$ SUSY QM algebra can be built based on
(\ref{eqmotion43dgf}), by using the matrix:
\begin{equation}\label{dfy}
\mathcal{D}=\left(%
\begin{array}{cc}
 2{\,}\partial_{w}+2{\,}\partial_{u} & -\frac{m^2}{2}\Big{(}\bar{w}(q_1+q_2)+\bar{u}(q_1-q_2)\Big{)} \\
  -m^2\Big{(}w(q_1+q_2)+u(q_1-q_2)\Big{)} & 2{\,}\bar{\partial}_{w}+2{\,}\bar{\partial}_{u} \\
\end{array}%
\right)
\end{equation}
acting on :
\begin{equation}\label{we23e}
\left(%
\begin{array}{c}
  \psi_{\bar{w}}+\psi_{\bar{u}} \\
  \chi_{\bar{w}} \\
\end{array}%
\right)
\end{equation}
It is obvious that, by using (\ref{dfy}), we can construct the
matrix $D^\dag$ and the rest of the SUSY algebra, such as the
Hamiltonian and so on.

In order the equations of motion (\ref{eqmotion3}) to be identical
to (\ref{eqmotion43dgf}) the following condition must be imposed
on the fields $\psi_{\bar{u}}$ and $\psi_{\bar{w}}$,
\begin{equation}\label{conalg}
\partial_{w}\psi_{\bar{w}}=\partial_{u}\psi_{\bar{u}}
\end{equation}
The implications of the above condition to the case of the three
matter curves are quite interesting. For the $\Sigma_1$ ($z_1=0$)
matter curve, since $\psi_{\bar{2}}=0$, relation (\ref{conalg})
would imply $\partial_2\psi_{\bar{1}}=0$. The curve $\Sigma_2$
($z_2=0$) has localized solutions when $\psi_{\bar{1}}=0$, and in
conjunction with (\ref{conalg}) we get the condition
$\partial_1\psi_{\bar{2}}=0$. In the same vain, one has for the
$\Sigma_3$ ($z_1+z_2=0$) curve $\partial_w\psi_{\bar{w}}=0$. The
conditions $\partial_1\psi_{\bar{2}}=0$ and
$\partial_2\psi_{\bar{1}}=0$ imply that $\psi_1$ is a function
only of $z_1$ (thus has no $z_2$ dependence) and $\psi_2$ is a
function only of $z_2$. This in turn would imply that the
functions $f(z_2)$ and $g(z_1)$ defined in relations
(\ref{locali1}), (\ref{locali2}) and (\ref{locali3}) are constant
functions, that is, $f(z_2)=c_1$ and $g(z_1)=c_2$, with $c_1$ and
$c_2$ arbitrary constants. Furthermore, the condition
$\partial_w\psi_{\bar{w}}=0$, implies that $\psi_{\bar{w}}$ is a
constant function, say $\psi_{\bar{w}}=c_3$. We summarize:
\begin{align}\label{generalresults}
&
{\,}{\,}{\,}{\,}{\,}{\,}{\,}{\,}{\,}{\,}{\,}{\,}{\,}{\,}{\,}{\,}{\,}{\,}{\,}{\,}{\,}{\,}{\,}{\,}{\,}{\,}{\,}{\,}{\,}{\,}{\,}{\,}{\,}{\,}{\,}{\,}{\,}{\,}{\,}{\,}{\,}{\,}{\,}{\,}{\,}{\,}{\,}{\,}{\,}{\,}{\,}{\,}\mathrm{Matter
{\,}{\,}Curve}{\,}{\,}{\,}z_1=0 \rightarrow f(z_2)=c_1
\\ \notag &
\partial_{w}\psi_{\bar{w}}=\partial_{u}\psi_{\bar{u}}{\,}{\,}{\,}{\,}{\,}\Rightarrow{\,}{\,}\mathrm{Matter{\,}{\,}
Curve}{\,}{\,}{\,}z_2=0 \rightarrow {\,}{\,}g(z_1)=c_2
\\ \notag & {\,}{\,}{\,}{\,}{\,}{\,}{\,}{\,}{\,}{\,}{\,}{\,}{\,}{\,}{\,}{\,}{\,}{\,}{\,}{\,}{\,}{\,}{\,}{\,}{\,}{\,}{\,}{\,}{\,}{\,}{\,}{\,}{\,}{\,}{\,}{\,}{\,}{\,}{\,}{\,}{\,}{\,}{\,}{\,}{\,}{\,}{\,}{\,}{\,}{\,}{\,}{\,}\mathrm{Matter
{\,}{\,}Curve}{\,}{\,}{\,}z_1+z_2=0 \rightarrow \psi_{\bar{w}}=c_3
\end{align}
The three conditions we just presented, are very much related to
the calculation of Yukawa couplings, when we have constant Higgs
wave function and absence of non-constant fluxes \cite{f1}.

As we saw earlier, the Yukawa coupling, in terms of the three
matter wave functions reads:
\begin{equation}\label{yukawa13}
Y=M_*^4\int_S\mathrm{d}^2z_1\mathrm{d}^2z_2{\,}\psi_1{\,}\psi_2{\,}\phi
\end{equation}
In the presence of background fluxes, the Yukawa coupling is given
by the overlapping integral \cite{f69},
\begin{equation}\label{gaugeyukawas}
Y^{ij}\sim
\int_{S}z_1^{3-i}z_2^{3-j}e^{M_{k\bar{l}}z_k\bar{z}_l}\dot
f(z_1,z_2)
\end{equation}
with $f(z_1,z_2)$ containing the gaussian profiles of the
localized fermions along the matter curves. When the $M_{i,j}$ is
constant or zero, there is a $U(1)\times U(1)$ symmetry, under
which the coordinates are invariant,
\begin{equation}\label{coordinateinv}
z_1\rightarrow
e^{ia_1}z_1{\,}{\,}{\,}{\,}{\,}{\,}{\,}{\,}z_2\rightarrow
e^{ia_2}z_2
\end{equation}
In that case, all Yukawas other than the $Y_{33}$, vanish. These
gauge symmetries are broken when $M_{i,j}$ has a non trivial gauge
dependence, which happens when background fluxes are turned on.
This case is particularly interesting, since in this way a
hierarchical fermion Yukawa matrix is obtained but we shall not
pursuit these issues further.

\noindent In the absence of fluxes, the fermionic matter functions
are equal to:
\begin{align}\label{wf}
& \psi_1=f(z_2)e^{-q_1m^2|z_1|^2}
\\ & \notag\psi_2=g(z_1)e^{-q_2m^2|z_2|^2}
\end{align}

\noindent In the special case that $\phi=\mathrm{const}$, the wave
function of the Higgs field is unlocalized, which means that the
Higgs field lives in the bulk rather than localized on a matter
curve. In the absence of non-constant fluxes, we have
$f(z_2)=g(z_1)=1$ and it can be proved that in this case, the only
non-vanishing Yukawa is the $Y_{33}$. This coupling gives mass to
the heaviest lepton and quark generations, and is equal to,
\begin{equation}\label{yukawa1}
Y_{33}\sim
M_*^4\int_S\mathrm{d}^2z_1\mathrm{d}^2z_2{\,}e^{-q_1m^2|z_1|^2}{\,}e^{-q_2m^2|z_2|^2}.
\end{equation}
Consequently, we see that the conditions (\ref{generalresults}),
imposed by the $N=2$ SUSY QM of the system described by the
equations of motion (\ref{eqmotion43dgf}), are the same with the
conditions that the matter curves wave functions satisfy, in order
to built Yukawa couplings in the absence of non-constant fluxes
with non-localized Higgs. Indeed the two cases are identical when
$c_3=c_2=1$. This type of Yukawa couplings is usually found in
type IIB and F-theory compactified on non del-Pezzo surfaces
\cite{f1}.

\noindent Before closing this section we discuss an important
issue. Spacetime supersymmetry and supersymmetric quantum
mechanics are not the same, nevertheless the connection is
profound, since extended (with $N=4,6...$) supersymmetric quantum
mechanics models describe the dimensional reduction to one
(temporal) dimension of $N=2$ and $N=1$ Super-Yang Mills models
\cite{pashev}. A serious question rises at this point. By looking
the $N=2$ SUSY QM algebra supercharges (\ref{dfy}), one could
thing that it is intended to embed any intersection curve in the
same sort of $N=2$ SUSY QM algebra. In general, a sort of $N=2$
supersymmetry is rather unexpected, since the intersection of
three D7-branes, breaks four-dimensional supersymmetry down to
$N=1$, and so the $N=2$ structure mentioned above is lost.
However, this is not true, since supersymmetry and supersymmetric
quantum mechanics is not the same. Indeed, the $N=2$ SUSY QM
supercharges do not generate spacetime supersymmetry. By the same
token, the supersymmetry in supersymmetric quantum mechanics, does
not relate fermions and bosons. The SUSY QM supercharges do not
generate transformations between fermions and bosons. These
supercharges generate transformations between two orthogonal
eigenstates of a Hamiltonian, eigenstates that are classified
according to their Witten parity. Hence, this flow of the $N=2$
breaking argument is not true, due to the non-spacetime structure
of SUSY QM.

\section{Global $U(1)$ Symmetries Along Matter Curves, Yukawa Couplings, Proton Decay Operators and Neutrino Mass Operators}

The $N=2$ supersymmetric quantum mechanics algebra is invariant
under an R-symmetry. Likewise, the Hamiltonian is also invariant
under this symmetry \cite{susyqm}. Actually, the superalgebra
(\ref{susy1}) and (\ref{susy3}) is invariant under the
transformation,
\begin{equation}\label{rsymetry}
\left (\begin{array}{c}
  Q'_1 \\
   Q'_2  \\
\end{array}\right )=\left (\begin{array}{ccc}
  \cos a & \sin a \\
  -\sin a & \cos a  \\
\end{array}\right ) \cdot \left ( \begin{array}{c}
  Q_1 \\
  Q_2 \\
\end{array} \right )
\end{equation}
with $a$ an arbitrary constant. Furthermore, the complex
supercharges $Q$ and $Q^{\dag}$ are transformed under a global
$U(1)$ transformation:
\begin{equation}\label{supercharges}
Q^{'}=e^{ia}Q, {\,}{\,}{\,}{\,}{\,}{\,}{\,}{\,}
{\,}{\,}Q^{'\dag}=e^{-ia}Q^{\dag}
\end{equation}
This R-symmetry is also a symmetry of the Hilbert states
corresponding to the subspaces $\mathcal{H}^{+}$ and
$\mathcal{H}^{-}$. Thus, the eigenfunctions of
$H_{+}=A{\,}A^{\dag}$ and $H_{-}=A^{\dag}{\,}A$, are invariant
under this $U(1)$-symmetry, namely,
\begin{equation}
|\psi^{'+}\rangle=e^{i\beta_{+}}|\psi^{+}\rangle,
{\,}{\,}{\,}{\,}{\,}{\,}{\,}{\,}
{\,}{\,}|\psi^{'-}\rangle=e^{i\beta_{-}}|\psi^{-}\rangle
\end{equation}
It is clear that the parameters $\beta_{+}$ and $\beta_{-}$ are
global parameters with $\beta_{+}\neq \beta_{-}$. Consistency with
relation (\ref{supercharges}) requires that
$a=\beta_{+}-\beta_{-}$. For our purposes we shall use only the
symmetry $|\psi^{'+}\rangle=e^{i\beta_{+}}|\psi^{+}\rangle$. The
implications of this symmetry are quite interesting, since this
implies that the localized fields on each matter curve are
invariant under this symmetry. Let us see this for the $z_1=0$
matter curve. Due to this $U(1)$ symmetry, the $Q$ and $Q^{\dag}$
supercharges of equation (\ref{wit2}) are invariant under the
transformation (\ref{supercharges}). Consequently the
eigenfunctions of $D_1$,
\begin{equation}\label{wee33}
\left(%
\begin{array}{c}
  \psi_{\bar{1}} \\
  \chi_1 \\
\end{array}%
\right)
\end{equation}
are invariant under the following transformation,
\begin{equation}\label{wee313}
\left(%
\begin{array}{c}
  \psi^{{\,}}_{ _{\bar{1}}} \\
  \chi_1^{{\,}} \\
\end{array}%
\right)'=e^{ib_{+}}\left(%
\begin{array}{c}
  \psi_{\bar{1}} \\
  \chi_1 \\
\end{array}%
\right)
\end{equation}
In the same way, the localized fields on the second fermion matter
curve $z_2$, are invariant under,
\begin{equation}\label{rsym2}
\left(%
\begin{array}{c}
  \psi^{{\,}}_{ _{\bar{2}}} \\
  \chi_2^{{\,}} \\
\end{array}%
\right)=e^{ib_{2+}}\left(%
\begin{array}{c}
  \psi_{\bar{2}} \\
  \chi_2 \\
\end{array}%
\right)'
\end{equation}
Finally, the localized fields on the Higgs matter curve
$z_1+z_2=0$, are invariant under,
\begin{equation}\label{rsym2}
\left(%
\begin{array}{c}
  \psi^{{\,}}_{ _{\bar{w}}} \\
  \chi_w^{{\,}} \\
\end{array}%
\right)'=e^{ib_{3+}}\left(%
\begin{array}{c}
  \psi_{\bar{w}} \\
  \chi_w \\
\end{array}%
\right)
\end{equation}
Note that the fields on each matter curve have the same
transformation properties. For convenience, we gather the results
in the following table,
\begin{center}\label{table223}
\begin{tabular}{|c|c|c|c|}
  \hline
 $U(1)$ &\bf{matter curve $z_1=0$} & \bf{matter curve $z_2=0$} & \bf{matter curve $z_1+z_2{\,}={\,}0$}  \\
  \hline
 $e^{ib_{+}}$ & $\chi_1$ and $\psi_{\bar{1}}$ & & \\
  \hline
$e^{ib_{2+}}$ &  & $\chi_2$ and $\psi_{\bar{2}}$ & \\
  \hline
 $e^{ib_{3+}}$ & &  & $\chi_w$ and $\psi_{\bar{w}}$ \\
  \hline
\end{tabular}
\\ \bigskip{ \bfseries{Table 2: $U(1)$-Classification of the Localized Fields on the Three Matter Curves}}
\end{center}
\bigskip

\noindent The localized fields on each matter curve are invariant
under this $U(1)$ symmetry presented above. However, certain
conditions must hold. Indeed, if this symmetry is an actual
symmetry of the localized fields, then the action (plus kinetic
terms) for the localized fields must be invariant under this
$U(1)$ symmetry. Let us examine for example the matter curve
$z_1=0$. The localized fields action reads (recall
$\eta=\psi_2=0$),
\begin{equation}\label{localizedfieldact}
I_{L}=\int_{R^{1,3}\times S}\mathrm{d}x^4\mathrm{Tr}\Big{(}\chi_1
\wedge\partial_A\psi_{\bar{1}}+\frac{1}{2}\psi_{\bar{1}} \wedge
[\varphi,\psi_{\bar{1}}]+\mathrm{h.c.}\Big{)}
\end{equation}
In addition the kinetic terms are of the form,
\begin{equation}\label{kinetic}
\int_{R^{1,3}\times
S}\psi_{\bar{1}}^{\dag}\psi_{\bar{1}}\mathrm{d}x^4,{\,}{\,}{\,}{\,}{\,}{\,}{\,}{\,}{\,}{\,}\int_{R^{1,3}\times
S}\chi_1^{\dag}\chi_1\mathrm{d}x^4
\end{equation}
It is obvious that under the $U(1)$ transformation,
\begin{equation}\label{transu1}
\chi'_1=e^{ib_{+}}\chi_1,
{\,}{\,}{\,}{\,}{\,}{\,}{\,}{\,}{\,}{\,}\psi'_{\bar{1}}=e^{ib_{+}}\psi_{\bar{1}}
\end{equation}
the kinetic terms (\ref{kinetic}) are invariant. Still, the action
(\ref{localizedfieldact}) cannot be invariant unless,
\begin{equation}\label{cond}
e^{2{\,}ib_{+}}=1
\end{equation}
Also, we suppose that the field $\phi$ is not affected by the
$U(1)$-symmetry. The condition (\ref{cond}) implies that
$b_{+}=\pi n$, with $n=0,1,2,...$

\noindent But the fields $\chi_1$ and $\phi$ belong to the same
susy multiplet, thus we would expect that this $U(1)$-symmetry
should be a symmetry of the whole action. It turns out that in
order the localized fields have this $U(1)$-symmetry, the $\phi$
field must not transform under this symmetry. The implications of
this symmetry are quite interesting, at least phenomenologically,
as we shall see.

\subsection{Proton Decay Operators, Dirac and Majorana Neutrino
Masses}

Let us recall how Yukawas are constructed within F-theory
\cite{vafa2,review4,f1,f3,f5,f13,f41}.

\noindent In F-theory, Yukawa couplings are considered to be
overlapping integrals of the three matter curves wave functions
over $S$. The matter curves are the two fermionic and the one
corresponding to the Higgs. Let the wave functions that describe
each fermionic matter curve be, $\psi_1$, $\psi_2$ describing
$\Sigma_1$ and $\Sigma_2$ respectively. Owing to the $N=1$
supersymmetry of the four dimensional theory, the wave function of
the Higgs curve, $\phi$, is the same with the function $\psi_w$,
corresponding to the $z_1+z_2=0$ curve, as we saw earlier. Then,
in terms of the three wave functions, the Yukawa coupling reads:
\begin{equation}\label{yukawa154}
Y=M_*^4\int_S\mathrm{d}^2z_1\mathrm{d}^2z_2{\,}\psi_1{\,}\psi_2{\,}\phi
\end{equation}
The Yukawa couplings give masses to fermions, therefore these
couplings are most welcome in the theoretical setup of the local
model. As we saw, each localized fermion field corresponding to
the two matter curves $z_1$ and $z_2$, obeys a SUSY QM
$U(1)$-symmetry, different for each matter curve (see table 2).
Yukawa couplings describe couplings between a quark and a
righthanded quark or between a lepton and a righthanded lepton. We
denote the lepton fields with the field operators $L=(N,E)$ and
also with $E^c$ the right handed one. Additionally, the quark
fields are represented by $Q=(U,D)$ and their righthanded
counterparts, $U^c,D^c$. The Yukawa's stem from the superpotential
and are of the form \cite{f1,f5},
\begin{equation}\label{}
W_{Yuk}=Y^{U}QU^cH^u+Y^{D}QD^{c}H^d+Y^{L}LE^{c}H^d
\end{equation}
We require the wave functions $\psi_1$ and $\psi_2$ in equation
(\ref{yukawa154}) to describe a lepton field and it's righthanded
field, or a quark field and it's righthanded field, respectively.
This means that leptons and quarks must be assigned to different
matter curves. This situation cannot be true in all local
geometrical GUT setups, like in the case of $SU(5)$, but can be
true in some cases, like in the flipped $SU(5)$ \cite{f39}
construction. Let the transformations of $\psi_1$ and $\psi_2$ be
that of table 2, that is,
\begin{equation}\label{ez}
\psi'_1=e^{ib_{+}}\psi_1,{\,}{\,}{\,}{\,}{\,}{\,}{\,}\psi'_2=e^{ib_{2+}}\psi_2
\end{equation}
Due to equation (\ref{cond}), the parameter $b_{+}$ is equal to
$b_{+}=\pi n$, with $n=0,1,2,...$ and similarly, $b_{1+}=\pi m$,
with $m=0,1,2,...$

\noindent In order the Yukawa coupling to be invariant under this
combined action of the $U(1)$'s, we easily find that the
parameters $b_{+}$ and $b_{2+}$ must be related as follows,
\begin{equation}\label{parametersrelation}
b_{+}=-b_{2+}
\end{equation}
On that account, we conclude that fermions belonging to a quark or
lepton family and their righthanded fermions (corresponding to
different matter curves), must have opposite transformation
properties under the SUSY QM-$U(1)$ symmetry, in order the Yukawa
couplings are invariant under this symmetry. Note that this
$U(1)$-symmetry is not a result coming from the local geometric
features of the surface $S$. It comes from the SUSY QM algebra
that the localized fields obey. The outcomes of the two conditions
(\ref{cond}) and (\ref{parametersrelation}), are quite interesting
phenomenologically. We consider first proton decay operators. The
proton decay operators are unwanted terms coming from the action.
The 4-dimensional proton decay operators are,
\begin{equation}\label{prot4}
W_{4a}\sim LLE^c{\,}{\,}{\,}{\,}{\,}W_{4b}\sim
QD^cL,{\,}{\,}{\,}{\,}{\,}W_{4c}\sim U^cD^cD^c
\end{equation}
Moreover, the 5-dimensional proton decay operators are,
\begin{equation}\label{prot5}
W_{5a}\sim \frac{1}{M}QQQL{\,}{\,}{\,}{\,}{\,}W_{5b}\sim
\frac{1}{M}LLH_uH_u,{\,}{\,}{\,}{\,}{\,}W_{5c}\sim
\frac{1}{M}U^cU^cD^cE^c
\end{equation}
The SUSY QM $U(1)$-symmetry restricts the proton decay operators,
as is obvious by looking the constraints (\ref{cond}) and
(\ref{parametersrelation}). Let us see which operators are allowed
subject to the SUSY QM $U(1)$-symmetry in detail. We study first
the dimension-4 operators. The operator $W_{4a}$ (see relation
(\ref{prot4})) is not allowed since although the  $LL$ part is
invariant (same fermions, see (\ref{cond})), the $E^c$ gives a
total $e^{ia}$ factor to the term. Likewise, $W_{4b}$ is not
invariant, since, although the $QD^c$ part is invariant (fermion
and corresponding righthanded fermion, see
\ref{parametersrelation})), the leptons $L$ have different
transformation properties from the quarks. The term $W_{4c}$ is
not allowed because, although the $D^cD^c$ is invariant, the $U^c$
gives an overall exponential factor to the term. Hence, the
dimension-4 proton decay operators of relation (\ref{prot4}) are
not allowed in the theory, if the SUSY QM $U(1)$-symmetry is
obeyed by the fermion fields localized on the matter curves.

\noindent Let us now check the dimension-5 operators. The operator
$W_{5a}$ is not invariant under the $U(1)$. Indeed, although the
$QQ$ part is invariant (same fermions) the $QL$ part is not
invariant since the first fermion is a quark and the second is a
lepton. For the same reasoning the operator $W_{5c}$ is not
invariant under the $U(1)$ so it cannot appear in the action. On
the contrary, the operator $W_{5b}$ is invariant, and thus can
affect the phenomenological outcomes off the model.

\subsection{Neutrino Masses and SUSY QM $U(1)$-Invariance}

\noindent The minimal $SU(5)$ F-theory GUT predicts Dirac and
Majorana neutrino masses \cite{f69}. Indeed, by integrating out
massive Kaluza-Klein modes, generates higher dimensional operators
that give phenomenologically acceptable masses for neutrinos.
Particularly, the Majorana mass F-term is of the form \cite{f69},
\begin{equation}\label{majorana}
\int\mathrm{d}\theta^2\frac{H_uLH_uL}{\Lambda_{UV}}
\end{equation}
When the Higgs field develops a vacuum expectation value, the
above term yields a Majorana mass for the neutrinos. The Majorana
mass term (\ref{majorana}) is clearly invariant under the SUSY QM
$U(1)$ symmetry because the term $LL$ is invariant (same fermions)
and the Higgs fields are not affected at all.

\noindent In the Dirac scenario, the D-term, generated by
integrating out massive Kaluza-Klein modes on the Higgs curves,
\begin{equation}\label{Dirac}
\int\mathrm{d}\theta^4\frac{H_d^{\dag}LN_R}{\Lambda_{UV}}
\end{equation}
gives a Dirac mass to the neutrinos. The field $N_R$ describes the
right-handed neutrino. The peculiarity of $N_R$ is due to that,
the right handed neutrino localizes on curves normal to the
GUT-seven brane \cite{f69}, a fact that put's in question the
local description concept of F-theory GUTs. Still, normal curves
can form part of a consistent local model \cite{f69}. The Dirac
mass term (\ref{Dirac}) is not invariant under the SUSY QM $U(1)$,
since from the term $H_d^{\dag}LN_R$, only the field $L$ is
transformed under the $U(1)$. Thus we can see that only the
Majorana mass terms is favored in the scenario we presented.

\section{Conclusions}

In this article we found that the fields localized at $D7$ branes
intersections are closely connected to an $N=2$ SUSY QM algebra.
Particularly, each matter curve corresponds to a different algebra
and due to this algebra, a global $U(1)$-symmetry underlies the
system. In view of this symmetry, the localized fields on each
matter curve satisfy certain conditions which we classified in
Table 2. Furthermore, since the Yukawa couplings are important to
GUT phenomenology, they must be invariant under this $U(1)$. This
condition, in conjunction with the table 2 transformations,
classifies the fermion transformations as in the following table,
\begin{center}\label{tab789}
\begin{tabular}{|c|c|c|c|c|c|}
  \hline
 \bfseries{Term} & $e^{ib_{2+}}$ & $e^{ib_{+}}$ & $e^{ia_{2+}}$ & $e^{ia_{+}}$ & $U(1)-$Invariant  \\
  \hline
 L & & &  & $e^{ia_{+}}$ & $U(1)-$Invariant\\
  \hline
LL  & &  &  &  & $U(1)-$Invariant\\
  \hline
$E^c$  & &  & $e^{ia_{2+}}$ &  &  \\
 \hline
Q &  & $e^{ib_{+}}$ &  & &  \\
 \hline
QQ &  & &  &  & $U(1)-$Invariant \\
 \hline
$D^c$ & $e^{ib_{2+}}$ &  & &  &  \\
 \hline
$U^c$ & $e^{ib_{2+}}$ &  & &  &  \\
 \hline
L$E^c$ &  & &  &  & $U(1)-$Invariant \\
 \hline
Q$D^c$ & &  & &  & $U(1)-$Invariant \\
 \hline
Q$U^c$ & &  & &  & $U(1)-$Invariant \\
 \hline
$U^cU^c$ & &  &  & &  $U(1)-$Invariant\\
 \hline
$D^cD^c$ & &  & &  & $U(1)-$Invariant \\
 \hline
\end{tabular}
\\ \bigskip{ \bfseries{Table 3: $U(1)$-Classification of various terms}}
\end{center}
Owing to the above transformation properties, we found
restrictions on the proton decay operators, many of which are not
allowed. Moreover, this $U(1)$ SUSY QM symmetry restricts the
neutrino mass operators. Particularly we found that only the
Majorana mass terms are allowed in our scenario.

\noindent We must mention that there are much more elaborated and
geometry inspired techniques to restrict proton decay operators
(see for example \cite{review1,review2,review3,review4,f9,f12}),
such as monodromies, but we do not discuss these here.

 Moreover, the requirement of an $N=2$ SUSY QM on a special
system leads to specific conditions on the fermion fields and (due
to supersymmetry) on the Higgs boson. As we found, these
conditions are met in the construction of the Yukawa coupling that
gives mass to the top quark, with a delocalized bulk Higgs.

In order to obtain the correct hierarchies and mixing of the
matter fields, external non-trivial background fluxes must be
turned on. We examined if the $N=2$ SUSY QM algebra still
underlies the localized fermionic solutions. We studied the
constant flux case and we found that the SUSY algebra still holds
for the two matter curves, namely $z_1=0$ and $z_2=0$. However,
for the Higgs curve, namely $z_1+z_2=0$, things are different. It
seems that the algebra is not a $N=2$ without central charge SUSY
QM algebra. We hope to address this problem in the future, but it
kind of surprised us. The surprise is due to the fact that the
adjoint vacuum expectation value $<\phi>$ remains the same as in
the flux-less case, so we did not expect things to change so
drastically.

\noindent Finally we performed a holomorphic perturbation of the
metric that describes the complex surface $S$ and we studied how
the perturbation modifies the net number of the zero modes that
the un-perturbed system has. We found that, due to a theorem
characteristic for Fredholm operators, the operators that describe
the perturbed and un-perturbed systems have equal indices. We
checked the validity of the theorem, for every matter curve and
Higgs curve. Unfortunately, this theorem does not gives us
information on the specific form of the wave functions.

\section*{Acknowledgments}

V. Oikonomou is indebted to Prof. G. Leontaris for useful
discussions on F-theory GUTs and related issues. Also the author
would like to thank the anonymous referee of NPB who's questions
and comments motivated section 5 of this article.

\end{document}